\documentclass{elsart1p}

\usepackage{graphicx}
\usepackage{amssymb}

\begin{document}

\begin{frontmatter}



\title{What the inflaton might tell us about RHIC/LHC}


\author{J{\"u}rgen Berges}

\address{Institute for Nuclear Physics, 
Darmstadt University of Technology,\\
Schlossgartenstr. 9, 64289 Darmstadt, Germany}

\begin{abstract}
Topical phenomena in high-energy physics related to collision experiments of heavy nuclei ("Little Bang") and early universe cosmology ("Big Bang") involve far-from-equilibrium dynamics described by quantum field theory. One example concerns the role of plasma instabilities for the process of thermalization in heavy-ion collisions. The reheating of the early universe after inflation may exhibit rather similar phenomena following a tachyonic or parametric resonance instability. Certain universal aspects associated to nonthermal fixed points even quantitatively agree, and considering these phenomena from a common perspective can be fruitful. 
\end{abstract}

%\begin{keyword}

%\PACS
%\end{keyword}
\end{frontmatter}

% main text
\section{Introduction}
\label{sec:intro}

Collision experiments of heavy nuclei 
at the Relativistic Heavy Ion Collider (BNL), the Large Hadron Collider (CERN) and future experiments at the FAIR facility (GSI) involve far-from-equilibrium dynamics
for strongly interacting matter described by quantum chromodynamics 
(QCD). Available data from RHIC reveals remarkable, unexpected properties such 
as rapid apparent thermalization with robust collective phenomena. There seems to be a good description of a wide range of data using hydrodynamical models with near-perfect fluidity, provided one assumes that suitable initial conditions form in about $0.5$--$2$ fm/c~\cite{Hydro,RR07}. The explanation of these findings from QCD provides a challenge for theory. Strong interactions can play an important role for QCD at accessible energies. However, there are theoretical indications that essential nonequilibrium dynamics could affect the picture dramatically. Complete local thermal equilibrium may not be necessary for the application of hydrodynamics~\cite{BBW04,ALMY05}. For sufficiently high energy densities nonequilibrium instabilities in anisotropic plasmas have been identified as the parametrically fastest processes, which may help to explain a rapid isotropization of the equation of state relevant for near-perfect fluid descriptions~\cite{Plasmainst,ALMY05,WongYangMills,Romatschke:2006nk,Berges:2007re}. 

Nonequilibrium instabilities may also play an important role for the process of thermalization in the early universe after inflation~\cite{preheat1,preheat2,preheat3,preheat4,Berges:2002cz,Arrizabalaga:2004iw}. In inflationary cosmology, the universe at early times expands quasi-exponentially in a vacuum-like state. During this stage of inflation, all energy is contained in a slowly evolving inflaton field. Eventually the inflaton field decays and transfers all of its energy to particles, thereby starting the thermal history of the hot Friedmann cosmology. In chaotic inflationary models for a wide range of couplings the particle production from a coherently oscillating inflaton occurs in the nonperturbative regime of a parametric-resonance instability. This picture, with variation in its details, extends to other inflationary models. 

Though heavy-ion collisions and inflationary cosmology involve energy scales many orders of
magnitude apart, their description requires similar quantum field theoretical techniques and 
certain universal aspects can even quantitatively agree. Instabilities lead to exponential growth of occupation numbers in long wavelength modes on time scales much shorter than the asymptotic thermal equilibration time. Though the underlying mechanisms are very different for QCD and for scalar inflaton dynamics, the subsequent evolution after an instability follows similar patterns: After a fast initial period of exponential growth the dynamics slows down
considerably. The system is still far from equilibrium at this stage and the subsequent slow evolution is characterized by power-law distributions. A new class of infrared scaling solutions, where a characteristic large correlation length leads to independence of long-distance properties from details of the underlying microscopic theory, has recently been found in the context of early-universe reheating dynamics~\cite{Berges:2008wm}. The signatures can be dramatic.  Nonthermal scaling solutions can exhibit strongly enhanced low-momentum fluctuations $\sim p^{-4}$, as compared to a thermal high-temperature distribution $\sim p^{-1}$. This long-distance behavior is distinct from scaling associated to (weak) turbulence, such as $\sim p^{-3/2}$, which is relevant for characteristic properties at shorter distances~\cite{Micha:2002ey}. It is remarkable that the same high-momentum scaling exponents known from inflaton dynamics have recently also been found to describe gluonic properties within lattice QCD in the classical-statistical gauge theory limit~\cite{BSS2}. 
Research concerning quantitative agreements of far-from-equilibrium properties in heavy-ion collision experiments and early universe dynamics is still in its infancies. This article highlights some known aspects and important open questions.

\section{Collisions of heavy nuclei and QCD plasma instabilities}

A wide range of experimental RHIC data, including elliptic flow measurements for momenta of less than about $1$-$2$ GeV, can be described using hydrodynamical models with near-perfect fluidity~\cite{Hydro,RR07}. The essential assumption of ideal hydrodynamic descriptions is the presence of an almost constant equation of state $P=P(\epsilon)$, which relates energy density $\epsilon$ to pressure $P$ given by a nearly diagonal stress tensor
\begin{equation}
T_{ij} \simeq P \delta_{ij}
\label{eq:isotropy}
\end{equation} 
in the local fluid rest frame. This is needed to close the system of equations obtained from the conservation of the energy-momentum tensor. Relation (\ref{eq:isotropy}) is true for isotropic systems if typical relativistic excitations have random directions, even if their energy distribution is far from thermal. Let "$T$" denote the characteristic momentum of typical excitations out of equilibrium, which may be identified with the saturation scale $Q_s$ at time $Q_s^{-1}$ in the saturation scenario~\cite{sat}. The relevant dephasing time which is required to obtain an almost constant or "prethermalized" equation of state for isotropic systems is $t_{\rm pt} \sim \mathcal{O}(1/T)$~\cite{BBW04}. For sufficiently weakly coupled theories this time scale is much faster than the inverse relaxation rate of arbitrarily small departures from equilibrium given by
\begin{equation}
t_{\rm relax} \sim \mathcal{O}(1/g^4 T) \, ,
\label{eq:trelax}
\end{equation}
which is associated to large-angle scattering or near-collinear splitting processes among quarks and gluons in QCD with gauge coupling $g$~\cite{g4T}. For anisotropic systems this relaxation time would characterize the time for isotropization in the absence of nonequilibrium instabilities \cite{Berges:2005ai}. However, in QCD nonabelian Weibel instabilities can drive local isotropization on a time scale, which is faster than ordinary perturbative scattering processes~\cite{ALMY05}. For sufficiently weak coupling and order-one anisotropy the inverse instability growth rate
\begin{equation}
t_{\rm iso} \sim \mathcal{O}(1/g T)
\end{equation}
is parametrically faster than the inverse relaxation rate (\ref{eq:trelax})~\cite{Plasmainst}. Consequently, understanding when a hydrodynamic description can first provide a good approximation is the same question as understanding the fastest characteristic process driving isotropization.  
  
Most quantitative estimates of these processes rely on a separation of scales between suitably defined 'soft' and 'hard' momenta for sufficiently small characteristic running gauge coupling~\cite{Plasmainst}. 
In this case the gluon degrees of freedom are described as classical fields for soft and classical particles for hard scales. The respective hard-loop effective theory of soft excitations is a nonabelian version of the linearized Vlasov equations of traditional plasma physics, which are based on collisionless kinetic theory for hard particles coupled to a soft classical field.\footnote{For numerical simulations which take into account the backreaction of the soft fields on the hard particles using a Boltzmann-Vlasov treatment see \cite{WongYangMills}.} This approach neglects quantum corrections and may also be considered as an approximation of the classical-statistical field theory limit of the respective quantum gauge theory~\cite{Romatschke:2006nk,Berges:2007re}. Classical-statistical lattice gauge theory provides a quantitative description in the presence of sufficiently large energy density or occupation numbers per mode. The simulations are done by numerical integration of the classical lattice equations of motion and Monte Carlo sampling of initial conditions. In the following we concentrate on classical-statistical lattice gauge theory simulations of Ref.~\cite{Berges:2007re}. The different approximations agree to a large extent in their predictions for characteristic growth rates at early times. They apparently disagree, in particular, concerning subsequent evolutions to turbulent flow which is described below. 
   
Collisions of heavy nuclei leave behind a plasma of quarks and mostly
gluons in a nearly flat region of space because of Lorentz contraction along the beam- or $z$-axis. If the quarks and gluons stream freely into the surrounding space, then the distribution of particles would become locally highly anisotropic. In this case the stress tensor quickly acquires an oblate form with $T_{xx} \sim T_{yy} \gg T_{zz}$. The fastest growing mode of plasma instabilities then has its wavevector along the normal direction and generates a prolate contribution to the stress, which pushes the system towards greater isotropy. 
Theoretical discussions of thermalization typically start from the saturation picture of high-energy heavy-ion collisions, where there is initially a nonperturbatively large phase-space
density $\sim 1/g^2$ of gluons with momentum of order the saturation scale $Q_s$~\cite{sat}. To be specific, we consider the extreme anisotropy case described by an effectively $\delta(p_z)$-like initial distribution. 
\begin{figure}[t]
%\vspace*{3.ex}
\includegraphics[scale=0.55]{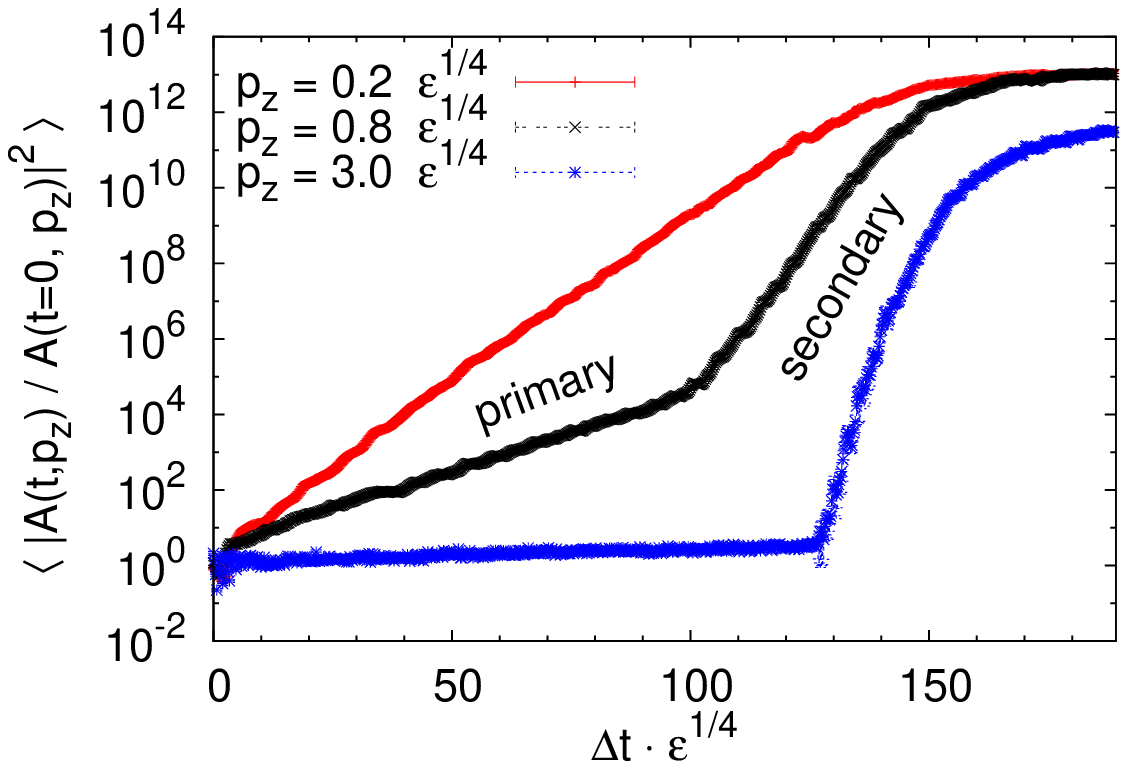}
\includegraphics[scale=0.5]{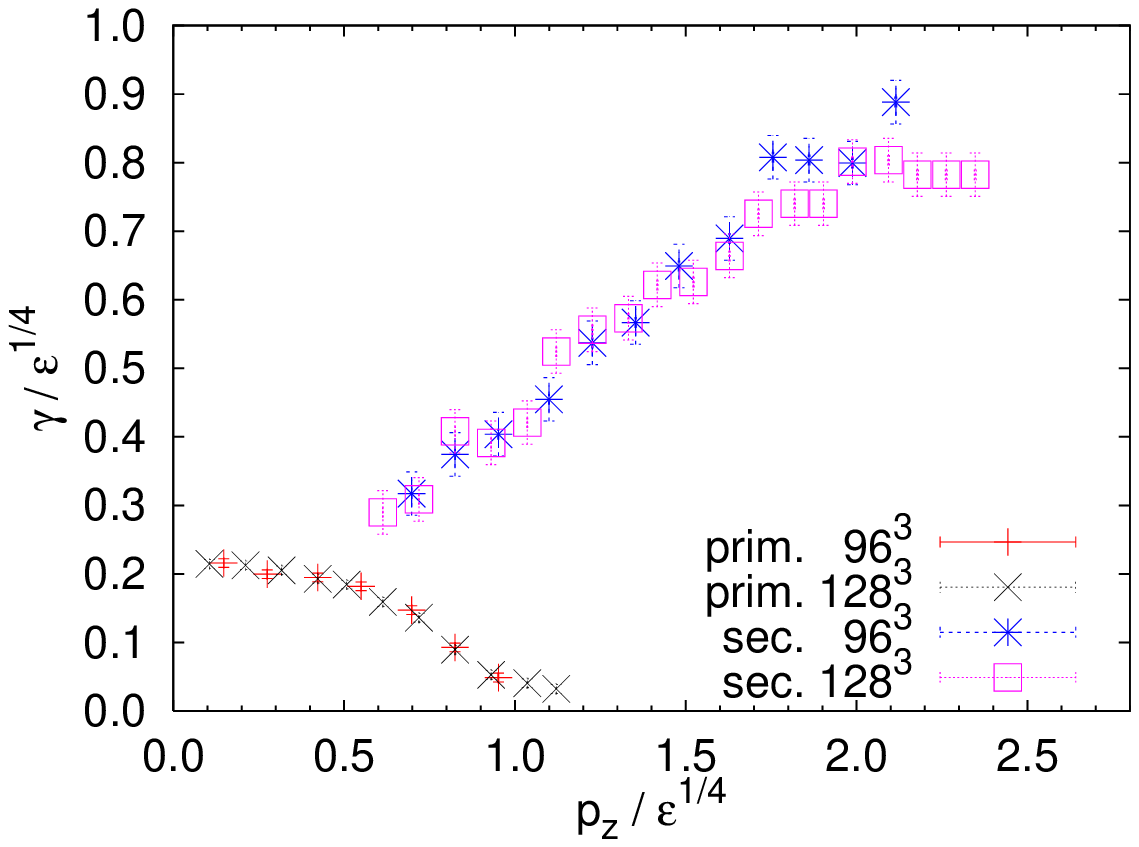}\caption{\label{fig:A-vs-t} LEFT: 
Fourier coefficients of the squared modulus of the gauge
field versus time for three different momenta parallel to the z-axis. The curves correspond to the momenta in the legend from top to bottom. RIGHT: Primary and secondary growth rates for $| A(t, p_z) |^2$ as a function of $p_z$ as measured on 96$^{3}$- and 128$^{3}$- lattices. (Taken from \cite{Berges:2007re}.)}
\end{figure}

The left graph of Fig.~\ref{fig:A-vs-t} shows the nonequilibrium time evolution of the color-averaged squared modulus of three different Fourier coefficients of the gauge field modes $A(t, \vec{p})$ for classical-statistical $SU(2)$ gauge theory in three spatial dimensions. They are displayed as a function of time, normalized by the corresponding field values at initial time.\footnote{The displayed results are for temporal-axial gauge. The characteristic growth rates have been verified also for Coulomb gauge~\cite{Berges:2007re}.} All values are given in appropriate units of the initial energy density $\epsilon$. Here $| A(t, \vec{p}) |^2$ may be associated to particle numbers divided by frequency. The plotted low-momentum modes clearly show exponential growth starting at the very beginning of the simulation. In contrast to these "primary" instabilities operating at low momenta, one observes from the left graph of Fig.~\ref{fig:A-vs-t} that momentum modes at sufficiently high momenta do not grow initially. The higher wave number modes typically exhibit exponential growth at a "secondary" stage that sets in later, but with a significantly larger growth rate. The right of Fig.~\ref{fig:A-vs-t} displays the momentum dependence of the growth rates for $| A(t, \vec{p}) |^2$ obtained from a fit to an exponential, which is done separately for the primary and secondary growth rates. The secondaries arise from fluctuation effects induced by the growth in the lower momentum modes, which can be seen by taking into account (2PI) resummed loop diagrams beyond the hard-loop approximation~\cite{Berges:2007re}. This growth saturates when all loop diagrams become of order one, which leads to a highly non-linear subsequent evolution. 

\begin{figure}[t]
%\vspace*{3.ex}
\includegraphics[scale=0.53]{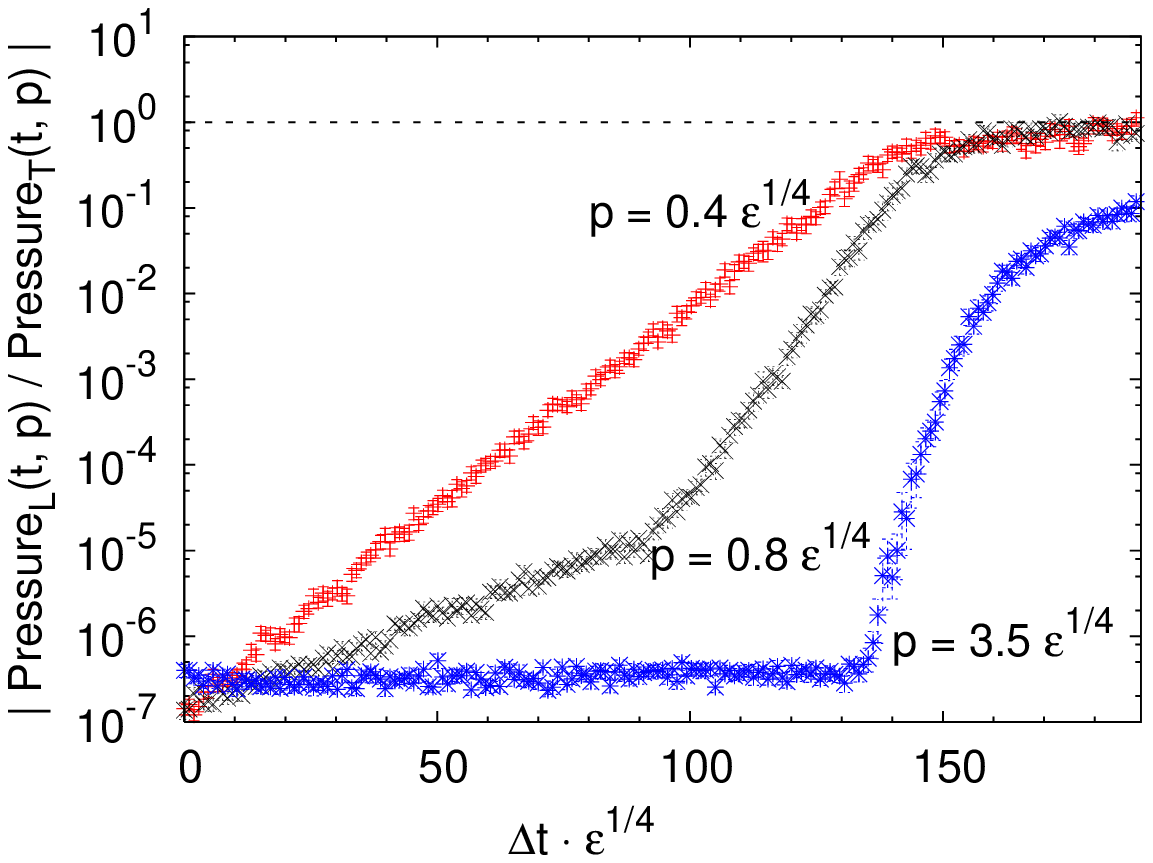}
\includegraphics[scale=0.53]{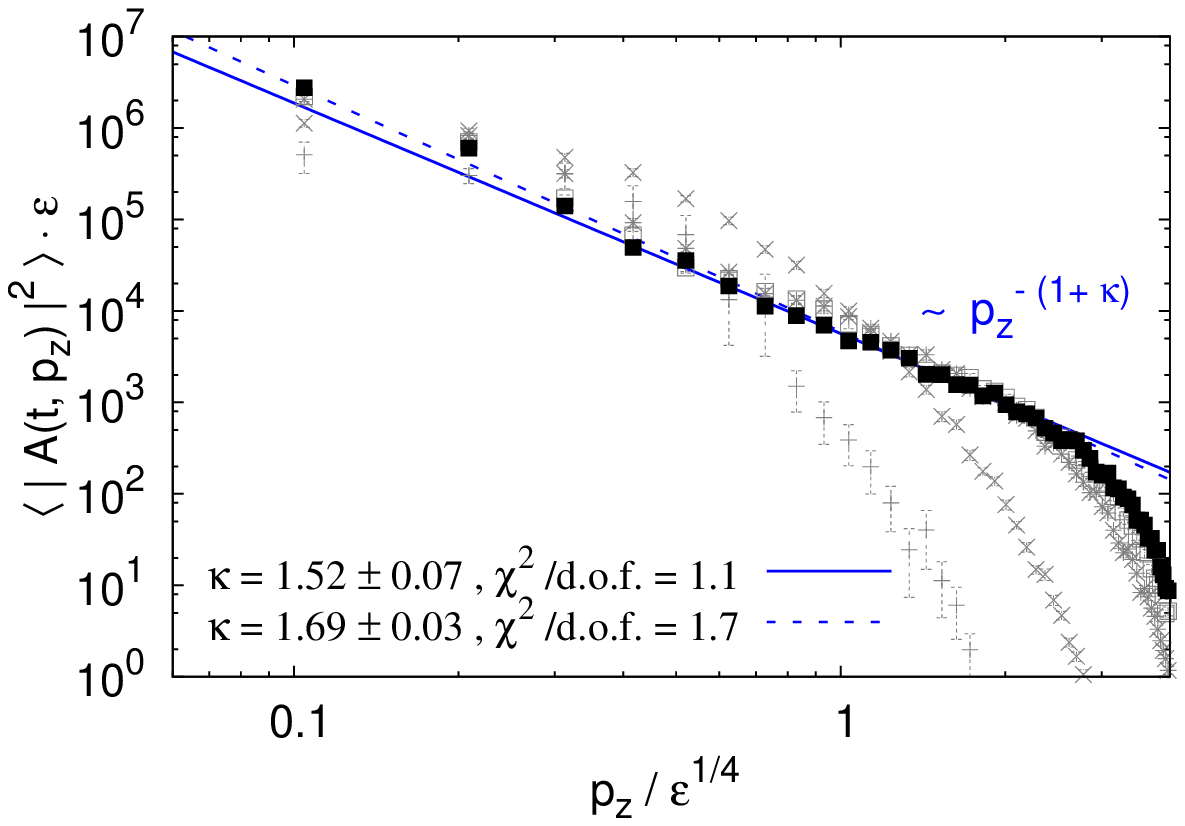}
\caption{\label{fig:pressure} LEFT: The ratio of modes of longitudinal pressure and transverse pressure as a function of time for different momenta~\cite{Berges:2007re}. RIGHT: Fourier coefficients of the squared modulus of the gauge field versus momentum for different snapshots of time. Fits to the spectrum at late time indicate a power-law with $\kappa = 3/2$ for the classical-statistical theory. The latest shown time is $t = 940 \epsilon^{-1/4}$ (black squares)~ \cite{BSS2}.}
\end{figure}

In order to obtain an estimate in physical units one may consider an initial (Bjorken) energy density of about $5$--$25$ GeV/fm$^3$ for RHIC experiments, and a projected factor of about two more for LHC energies. The inverse of the maximum primary growth rate for $|\, A(t, \vec{p}) \, |^2$ is $\gamma_{\rm max}^{-1} \simeq 4 \, \epsilon$, which then corresponds to a characteristic time scale of
\begin{equation}\label{eq:inverse-primary-gr}
 \gamma_{\rm max}^{-1}  \,\simeq\, 1.2 - 1.8 \, {\rm fm}/c \,\,\, {\rm (RHIC)}\,\,\, ,
 \qquad \gamma_{\rm max}^{-1}  \,\simeq\, 1.0 - 1.5 \, {\rm fm}/c \,\,\, {\rm (LHC)}.
\end{equation}
One observes that the results are rather insensitive to the precise value of the initial energy density because of its scaling with the fourth root of $\epsilon$. For comparison, a time scale associated with the largest observed secondary growth rates is about a factor of three shorter than what is given in (\ref{eq:inverse-primary-gr}). However, even though secondaries can reach considerably higher growth rates than primaries, they start later. As a consequence, a certain range of higher momentum modes can 'catch up' with initially faster growing infrared modes before the exponential growth stops, as seen in Fig.~\ref{fig:A-vs-t}. This leads to a fast effective isotropization of a finite momentum range, while higher momentum modes do not isotropize on a time scale characterized by plasma instabilities. An example of a gauge invariant quantity, which reflects this behavior, is given by the Fourier modes of the stress tensor $T_{ij}$. The left graph of Fig.~\ref{fig:pressure} shows the ratio of modes for longitudinal pressure $P_L(t,{\bf p}) \equiv T_{33}(t,{\bf p})$ with the corresponding transverse pressure for different absolute value of momentum. One observes a 'bottom-up isotropization' where for low-momentum
modes the ratio turns to one rather quickly, while high-momentum modes
only become isotropic at far later times. For the above energy densities one finds effective isotropization up to a characteristic momentum of about
\begin{eqnarray}
|{\bf p}| & \lesssim & 1 \, {\rm GeV} \, .
\end{eqnarray}
This is similar to the range of momenta for which hydrodynamic descriptions work, which typically show sizeable deviations from data above about $2$ GeV. 

The subsequent slow evolution, after the fast exponential growth stopped, seems to provide a suitable condition for hydrodynamic models as effective low-energy descriptions based on sufficiently small gradients. What is the physics underlying this slowing-down? Most investigations concentrate on power cascades towards shorter wavelengths reminiscent of Kolmogorov wave turbulence~\cite{Arnold:2005qs,Mueller:2006up,BSS2}. For the correlator of Fig.~\ref{fig:A-vs-t} such a behavior would be characterized by a power-law exponent $\kappa$, with
\begin{equation}
\langle |A(t,{\bf p})|^2 \rangle \sim |{\bf p}|^{-(1+\kappa)} \, .
\label{eq:powerlaw}
\end{equation}
The right graph of Fig.~\ref{fig:pressure} shows the result of classical-statistical simulations in Coulomb gauge for transverse fields and momenta ${\bf p}$ aligned to the $z$-direction. One observes that as time proceeds more and more ultraviolet modes approach a power-law, which is rather accurately described by $\kappa = 3/2$. For sufficiently high momenta this value may be compared to perturbative estimates. It has been pointed out that this value for $\kappa$ corresponds to the analytical result obtained from (2PI) resummed one-loop perturbation theory~\cite{BSS2}. Below we will see that the very same exponent is also found for inflaton dynamics. It has been noted that scaling behavior for gauge theories could deviate from scalar field dynamics and various estimates have been given for the value of $\kappa$ in the literature. In particular, the value two for $\kappa$ reported in Ref.~\cite{Arnold:2005qs} based on Vlasov equations seems excluded by the $\chi^2$-fit for the classical-statistical simulation of Fig.~\ref{fig:pressure}. The difference might be attributed to the fact that the hard modes in the Vlasov treatment represent static sources that lead to a stationary-state solution, which is not present due to total energy conservation in the classical-statistical lattice gauge theory simulation. Simulations on much larger lattices might be required to see whether the value two for $\kappa$ could characterize some intermediate-time behavior, which seems not the case in Fig.~\ref{fig:pressure}. Clearly, none of the classical approximations are sufficient to quantitatively address the late-time behavior, which should finally be characterized by a Bose-Einstein distribution for the gluons. While for low momenta the employed classical-statistical simulations are expected to give an accurate description for sufficiently high energy densities or occupation numbers, the high-momentum behavior will be altered by quantum corrections. It is remarkable that in the context of simpler scalar inflaton dynamics, where the {\em quantum} evolution can be addressed with present-day techniques, very similar phenomena are observed.

\section{Early universe reheating: A quantum example}

During inflation the universe exhibits a strongly accelerated expansion such that matter and radiation dilute very quickly. As a consequence, after inflation there is an enormous heating required in order to connect to the subsequent thermal history of the hot Friedmann cosmology. In a large variety of inflaton models nonequilibrium instabilities play a crucial role for the process of thermalization. For instance, in chaotic inflationary models a parametric-resonance instability leads to an exponential growth of occupation numbers~\cite{preheat1,preheat2,Berges:2002cz}. In hybrid inflation, including $D$-term inflation, inflaton decay proceeds via a tachyonic (spinodal) instability of the inhomogeneous modes which accompany symmetry breaking~\cite{preheat4,Arrizabalaga:2004iw}. Despite the different underlying physical mechanisms, the subsequent evolution in these models after an instability follows very similar patterns. 

In quantum field theory dynamics of parametric resonance~\cite{Berges:2002cz,Berges:2008wm} as well as tachyonic preheating~\cite{Arrizabalaga:2004iw} was studied in detail for $N$-component inflaton models using (2PI) resummed $1/N$ expansions to next-to-leading order (NLO)~\cite{Berges:2001fi,Aarts:2002dj}. These type of approximation are known to describe the late-time approach to thermal equilibrium characterized by Bose-Einstein~\cite{Berges:2000ur,Berges:2001fi} or Fermi-Dirac distributions~\cite{Berges:2002wr,BBW04}, respectively. 
Characteristic far-from-equilibrium phenomena,
such as an early prethermalization of the equation of state, have
been quantitatively studied in that context~\cite{BBW04,Dufaux:2006ee}. For a review see Ref.~\cite{Berges:2004yj}. 
\begin{figure}[t]
%\vspace*{3.ex}
\includegraphics[scale=0.45]{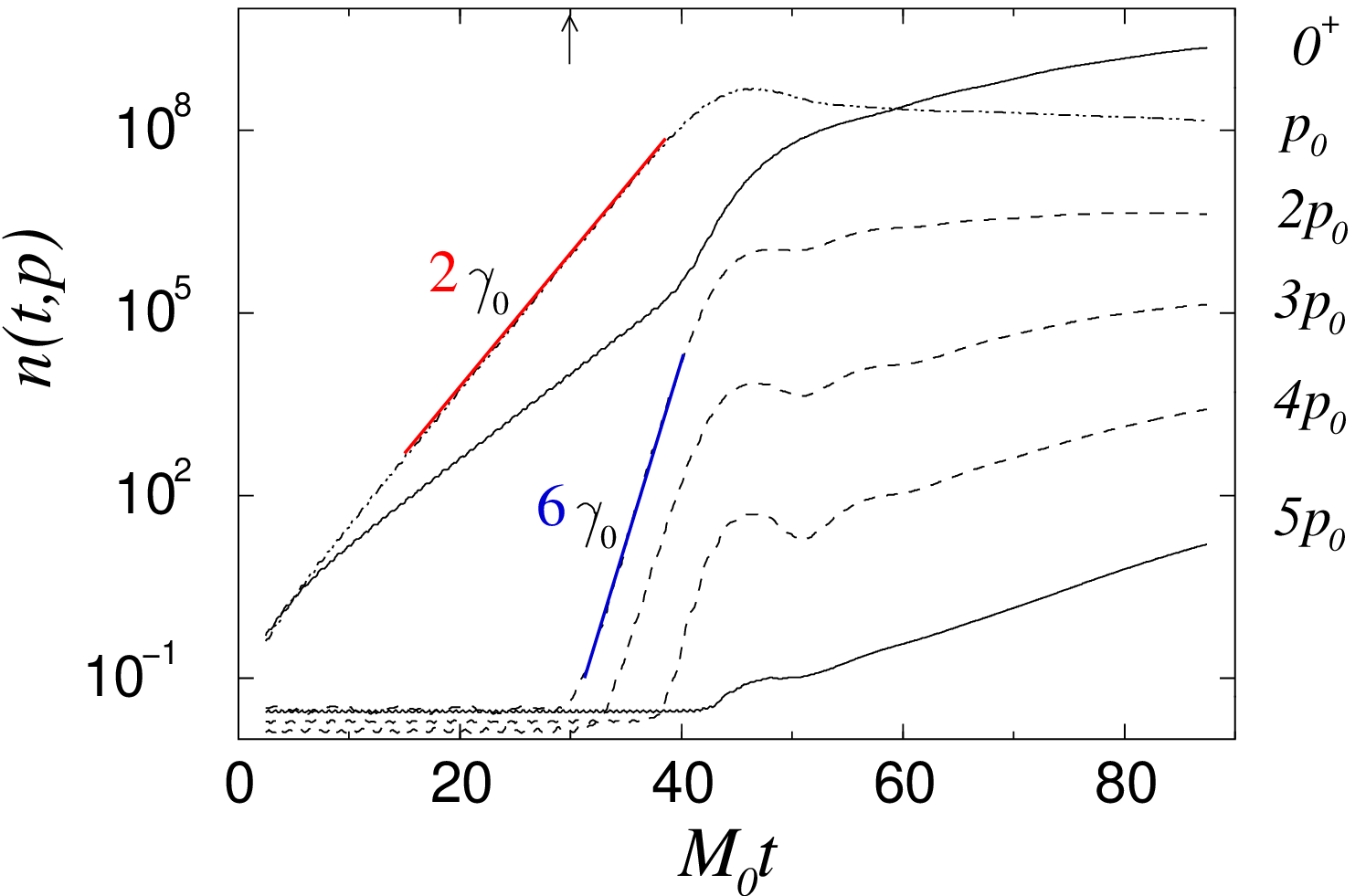}
\includegraphics[scale=0.26]{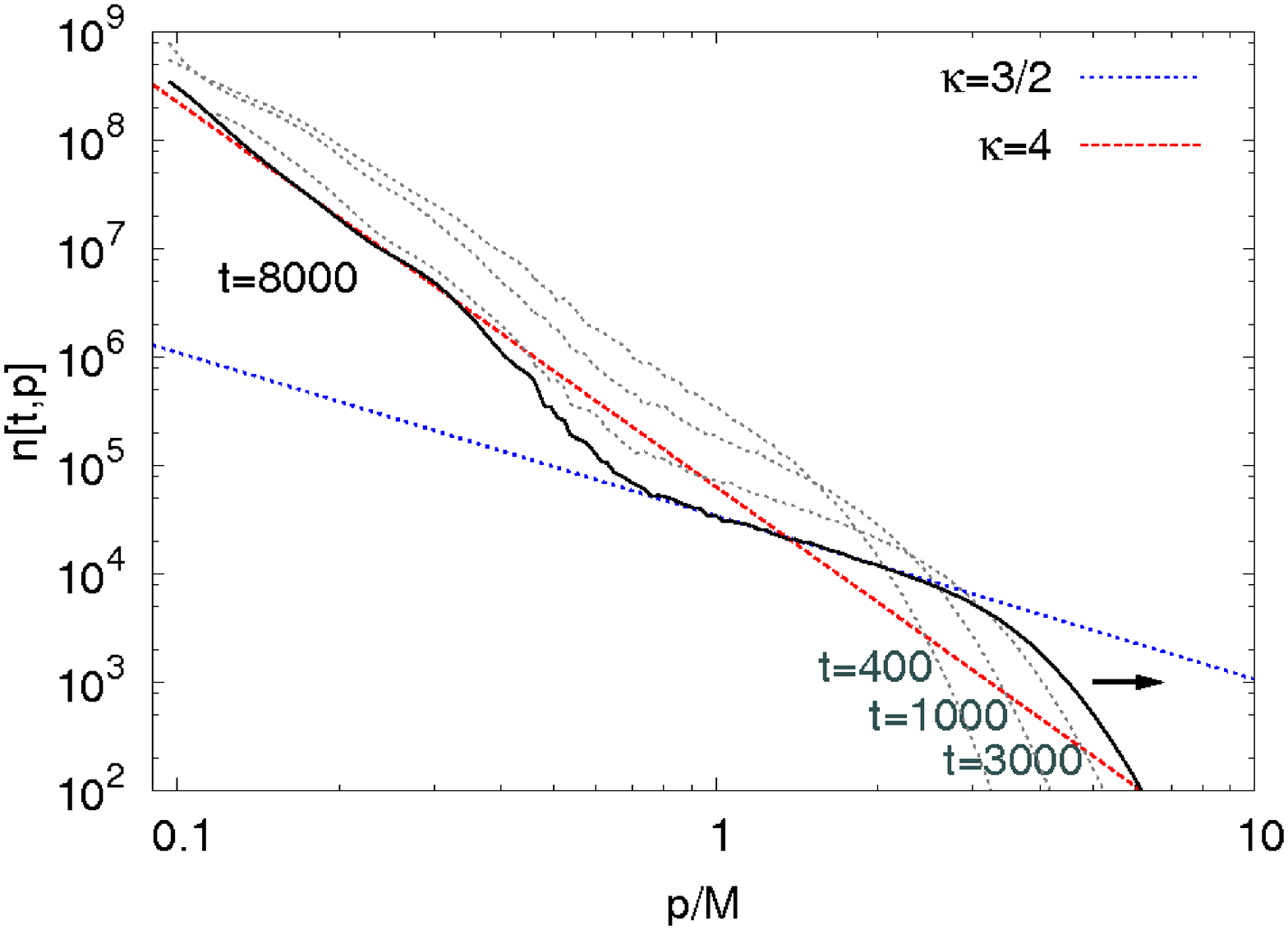}
\caption{\label{fig:inflaton} LEFT: Exponential early-time behavior for the inflaton occupation number as a function of time for various momenta following a parametric-resonance instability~\cite{Berges:2002cz}. RIGHT: Occupation number distribution for several later instances of time. At low momenta the nonperturbative critical behavior is rather well described by the infrared exponent $\kappa = 4$ (dashed line). For higher momenta the perturbative exponent $\kappa = 3/2$ (dotted line) related to Kolmogorov wave turbulence is approached. (From~\cite{Berges:2008wm}.)}
\end{figure}

A characteristic example for chaotic inflation models is given in Fig.~\ref{fig:inflaton}, which shows results for parametric-resonance amplification of occupation numbers in a weakly coupled $O(4)$-symmetric theory with quartic self-coupling $\lambda$~\cite{Berges:2002cz,Berges:2008wm}. The "occupation number" modes are defined as $n(t,{\bf p}) + 1/2 = \left( \langle |\varphi(t,{\bf p})|^2 \rangle_{\rm c} \langle |\dot{\varphi}(t,{\bf p})|^2\rangle_{\rm c} \right)^{1/2}$ using connected correlation functions for the spatial Fourier modes  $\varphi(t,{\bf p})$ of the scalar inflaton field. The left graph shows the characteristic exponential early-time behavior of various momentum modes in units of the mass scale set by the rescaled initial inflaton field amplitude $\sim \langle \varphi(t=0,{\bf x}) \rangle \sqrt{\lambda/6N}$. It is instructive to compare the initially fastest growing mode with momentum $p_0$ and primary growth rate $2 \gamma_0$ with the behavior at a higher momentum $2 p_0$. The latter does not grow at first but exhibits exponential growth at a secondary stage with the larger growth rate $6 \gamma_0$, i.e.\
multiples of the primary growth rate. This occurs in the non-linear regime, where parametrically $\langle |\varphi(t,{\bf p})|^2 \rangle \sim {\cal O} (N^{1/3}\lambda^{-2/3})$. Here occupied low-momentum modes act as sources for the secondary stage of enhanced amplification in a higher momentum range. The exponential growth stops when 
$\langle |\varphi(t,{\bf p})|^2 \rangle \sim {\cal O} (N^0\, \lambda^{-1})$ and the dynamics slows down considerably. At this nonperturbative stage all processes are of order unity and the system is strongly correlated despite the underlying weak coupling. It is striking to compare to the qualitatively similar primary and secondary growth stages for gauge fields in the context of plasma instabilities in Fig.~\ref{fig:A-vs-t}. 

At later times the comparison between gluon and inflaton dynamics shows certain characteristic properties, which even quantitatively agree. The occupation number distribution for the inflaton is displayed for several instances of time in the right graph of Fig.~\ref{fig:inflaton} \cite{Berges:2008wm}. The evolution is characterized by critical slowing down and the quasi-stationary solid curve exhibits different power-law regimes. At higher momenta the evolution approaches $n(t,{\bf p}) \sim |{\bf p}|^{-\kappa}$ with $\kappa = 3/2$, which is indicated by the corresponding dotted line in Fig.~\ref{fig:inflaton}. This is precisely the value characterizing the power-law behavior of gauge field fluctuations in the right graph of Fig.~\ref{fig:pressure}. The behavior is associated to weak Kolmogorov wave turbulence and can be obtained in perturbation theory. It should be emphasized that quantum corrections obstruct a power-law at very high momenta for which occupation numbers are of order one.

The situation changes dramatically in the infrared below a momentum scale at which occupation numbers grow nonperturbatively large to $n(t,{\bf p}) \sim 24 \pi/\lambda$ for $\lambda \ll 1$. A power-law behavior with strongly enhanced fluctuations emerges at low momenta according to the right graph of Fig.~\ref{fig:inflaton}. This is indicated by the dashed line, which corresponds to a critical exponent $\kappa = 4$. Analytically one finds that this is described by a nonthermal infrared fixed point solution of the renormalization group equations for the underlying quantum field theory, which is in the same universality class as the corresponding classical-statistical theory~\cite{Berges:2008wm,Berges:2008sr}. It is still an open question whether gauge field theories exhibit this new type of infrared fixed points. The corresponding right graph of Fig.~\ref{fig:pressure} for gauge field fluctuations in Coulomb gauge might indicate the slow emergence of a different infrared regime~\cite{BSS2}, and other indications seem to exist from reconsidering Vlasov results~\cite{Arnold:2005qs}. However, simulations for larger volumes and an ambitious analysis in terms of gauge invariant quantities such as stress-tensor correlation functions may be required to settle this important question.

\section{Conclusions}

Despite important differences the discussion of the thermalization process in heavy-ion collisions and cosmology after inflation shows remarkable similarities. In both cases nonequilibrium instabilities can lead to a fast period of exponential growth of occupation numbers. This is followed by a slow period, where the quantitative agreement concerns characteristic exponents or scaling functions describing scale-invariant properties of far-from-equilibrium dynamics. 

For a considerable class of inflaton models a detailed understanding in quantum field theory is available: Instabilities do not lead to fast thermalization, but they lead to an early prethermalization of bulk quantities such as the equation of state. The subsequent evolution approaches power-law behavior, which is described by nonthermal renormalization group fixed points of the underlying quantum theory. Similar to standard discussions in thermal equilibrium, one may distinguish between nonthermal infrared and ultraviolet fixed points. Important signatures of the nonperturbative infrared fixed points are related to strongly enhanced fluctuations on long-distance scales. This critical phenomenon is distinct from weak Kolmogorov wave turbulence relevant at shorter distances, which can be associated to the presence of an ultraviolet fixed point in the classical limit.

Nonthermal infrared fixed points have the dramatic consequence that a diverging time scale exists far from equilibrium, which can prevent or substantially delay thermalization. They are approached from nonequilibrium instabilities without fine-tuning of parameters. Estimates of the reheating temperature after inflation have to take this nonperturbative physics into account. A conservative limit requires thermal equilibrium at a reheating temperature of order $10$ MeV before Big Bang Nucleosynthesis. This can already rule out some simple weakly coupled inflaton models following parametric or tachyonic preheating dynamics, however, more realistic models have to be considered.
 
In the context of heavy-ion collisions, QCD plasma instabilities may lead to a regime with scaling behavior. Classical simulations indicate possible fast "bottom-up" isotropization due to instabilities with characteristic time scale of $1$-$2$ fm/c for low momenta of less than about a GeV. This appears to be near the range where hydrodynamic estimates give quantitative descriptions of experimental data. However, more refined calculations in expanding geometries including quantum corrections for high-momentum modes may be required to settle this question. It is remarkable that simulations in classical-statistical lattice gauge theory show quantitatively the same scaling exponents as for the inflaton in the turbulent regime at higher momenta. The question of whether it is possible to find a nonperturbative infrared fixed point in QCD in possibly the same universality class is exciting.   

\vspace*{0.2cm}

\noindent
This work is supported in part by the BMBF grant 06DA267, and by the
DFG under contract SFB634. Part of this work was done during the programme 
on "Nonequilibrium Dynamics in Particle Physics and
Cosmology" (2008) at the Kavli Institute for Theoretical Physics in Santa Barbara,
supported by the NSF under grant PHY05-51164.

\end{document}